\documentclass[aps,prl,twocolumn,superscriptaddress,amsmath,amssymb,showpacs]{revtex4}
\usepackage{color}
\usepackage{graphics}

\begin{document}

\title{Gravitation and regular Universe without dark energy and dark matter}%

\author{A.~V.~Minkevich}

\email{minkav@bsu.by, awm@matman.uwm.edu.pl}

\affiliation{Department of Theoretical Physics and Astrophysics, Belarussian
State University, Minsk, Belarus}

\affiliation{Department of Physics and Computer Methods, Warmia and Mazury
University in Olsztyn, Poland}


\begin{abstract}
It is shown that isotropic cosmology in the Riemann-Cartan spacetime allows to
solve the problem of cosmological singularity as well as the problems of
invisible matter components -- dark energy and dark matter. All cosmological
models filled with usual gravitating matter satisfying energy dominance
conditions are regular with respect to energy density, spacetime metrics and
the Hubble parameter. At asymptotics cosmological solutions of spatially flat
models describe accelerating Universe without dark energy and dark matter, and
quantitatively their behaviour is identical to that of standard cosmological
$\Lambda CDM$-model.
\end{abstract}


\maketitle

\section{Introduction}

The general relativity theory (GR) is the base of gravitation theory,
relativistic cosmology and astrophysics. At the same time GR leads to some
principal problems and difficulties, which in particular appear in cosmology.
The most principal problem of modern cosmology is connected with invisible
matter components in the Universe -- dark energy (DE) and dark matter (DM). The
explanation of observational cosmological data in the frame of standard
$\Lambda CDM$-model leads to conclusion that about 96\% energy in the Universe
is connected with DE and DM and only about 4\% energy is related to usual
barionic matter. As a result the present situation in cosmology and gravitation
theory is similar to that in physics at the beginning of XX century, when the
notion of "ether" was introduced with the purpose to explain various
electrodynamic phenomena. Another principal problem of standard cosmology,
which does not have acceptable solution still, is the problem of the beginning
of the Universe in time in the past -- the problem of cosmological singularity
(PCS).

Many attempts were undertaken with the purpose to solve indicated problems in
the frame of GR and existent candidates to quantum gravitation theory (string
theory/M-theory, loop quantum gravity) as well as of different generalizations
of GR (see for example \cite{1}). Although a number of results were obtained,
the search of the most satisfactory solutions of discussed problems is
continued. As it was shown in a number of papers (see \cite{2, 3, 4, 5, 6, 7,
8, 9} and Refs herein) the gravitation theory in 4-dimensional Riemann-Cartan
spacetime $U_4$ -- the Poncar\'e gauge theory of gravity (PGTG) -- offers
opportunities to solve indicated cosmological problems. Note at first, that the
PGTG is based on well known acceptable physical principles including gauge
invariance principle. Indeed the PGTG is a necessary generalization of metric
gravitation theory, if the Lorentz group is included to gravitation gauge
group, and namely the PGTG but not metric gravitation theory corresponds to
supergravity theory. The simplest PGTG is the Einstein-Cartan theory of gravity
\cite{10} based on gravitational Lagrangian in the form of scalar curvature of
spacetime $U_4$. The Einstein-Cartan theory leads to algebraic linear relation
between spacetime torsion and spin of gravitating matter, and generally  this
theory is considered as extension of GR to include spin momentum. However, if
we take into account that in the frame of PGTG the torsion tensor plays the
role of the gravitational gauge strength corresponding to spacetime
translations, which are connected directly with energy-momentum tensor, the
disappearance of the torsion in the case of spinless matter in the frame of
Einstein-Cartan theory testifies rather about degenerate character of this
theory \cite{1}. The situation comes to normal in the frame of PGTG based on
the gravitational Lagrangian ${\cal L}_{\rm g}$ including besides scalar
curvature similarly to Yang-Mills fields theory invariants quadratic with
respect to gravitational gauge field strengths -- the curvature and torsion
tensors \cite{11}. Because the quadratic part of ${\cal L}_{\rm g}$ is unknown,
the isotropic cosmology was built by using general expression of ${\cal L}_{\rm
g}$ including different invariants quadratic in the gravitational field
strengths. Let us to remind the most important physical results obtained in
Refs. 1-9 in the frame of isotropic cosmology in the Riemann-Cartan spacetime.
Any homogeneous isotropic model (HIM) in the frame of PGTG is described by
means of three functions of time -- the scale factor of Robertson-Walker
metrics $R$ and two torsion functions $S_1$ and $S_2$ determining non-vanishing
components of torsion tensor (unlike $S_1$ the torsion function $S_2$ is
pseudoscalar with respect to spatial inversions). Two types of HIM were built
and investigated: HIM with the only torsion function $S_1$ and HIM with two
torsion functions. Isotropic cosmology based on HIM of the first type allows to
solve the PCS \cite{2, 3, 4}: all cosmological models filled with usual matter
satisfying energy dominance conditions (including inflationary models) are
regular with respect to energy density, scale factor $R$ with its time
derivatives by virtue of gravitational repulsion effect at extreme conditions
in the beginning cosmological expansion provoked by spacetime torsion. However,
the situation with DE- and DM-problems in the case of these HIM becomes the
same as in GR. Isotropic cosmology based on HIM with two torsion functions
allows to solve the PCS as well as the DE-problem \cite{6, 8}. Moreover, the
DM-problem together with DE-problem can be solved in the frame of such HIM
\cite{7}. However, simultaneous solution of PCS and DE- with DM-problems was
not found still.

As it is shown in this paper all discussed problems can be solved in the frame
of isotropic cosmology in the Riemann-Cartan spacetime by certain restrictions
on indefinite parameters of gravitational Lagrangian ${\cal L}_{\rm g}$.

We will consider the PGTG based on sufficiently general following expression of
gravitational Lagrangian (definitions and notations of \cite{6, 9} are used
below):
\begin{eqnarray}
\label{1} {\cal L}_{\rm g}=\left[f_0\,
F+F^{\alpha\beta\mu\nu}\left(f_1\:F_{\alpha\beta\mu\nu}
                +f_2\: F_{\alpha\mu\beta\nu}+f_3\:F_{\mu\nu\alpha\beta}\right) \right. \nonumber \\ \left.
        + F^{\mu\nu}\left(f_4\:F_{\mu\nu}+f_5\: F_{\nu\mu}\right)+f_6\:F^2
    \right.  \nonumber
 \\ \left.
    +S^{\alpha\mu\nu}\left(a_1\:S_{\alpha\mu\nu}+a_2\: S_{\nu\mu\alpha}\right)
    +a_3\:S^\alpha{}_{\mu\alpha}S_\beta{}^{\mu\beta}\right].
\end{eqnarray}
The Lagrangian (1) includes the parameter $f_0=(16\pi G)^{-1}$ ($G$ is Newton's
gravitational constant, the light velocity $c=1$) and a number of indefinite
parameters: $f_i$ ($i=1,2,...6$) and $a_k$ ($k=1,2,3$). Physical consequences
of PGTG depend essentially on restrictions on indefinite parameters $f_i$ and
$a_k$. Some of such restrictions will be given below by investigation of HIM.

Gravitational equations for HIM with two torsion functions corresponding to
gravitational Lagrangian (1) allow to obtain cosmological equations
generalizing Friedmann cosmological equations of GR and equations for torsion
functions given in general form in \cite{9}. These equations contain five
indefinite parameters:
\begin{eqnarray}
  a = 2a_1  + a_2  + 3a_3, \qquad b = a_2  - a_1,
\nonumber\\
  f = f_1  + \frac{{f_2 }} {2} + f_3  + f_4  + f_5  + 3f_{6}\, ,
\nonumber\\
  q_1  = f_2  - 2f_3  + f_4  + f_5  + 6f_{6}, \qquad q_2  = 2f_1  - f_2,
\nonumber
\end{eqnarray}
and their mathematical structure and physical consequences depend essentially
on restrictions on these parameters. Unlike metric gravitation theory,
quadratic in the curvature terms of ${\cal L}_{\rm g}$ do not lead to higher
derivatives of $R$ in cosmological equations; higher derivatives can appear
because of terms of ${\cal L}_{\rm g}$ quadratic in the torsion tensor
\cite{11, 9}; in order to exclude higher derivatives of $R$ from cosmological
equations we have to put the restriction $a=0$ \footnote{Isotropic cosmology
with $a \neq 0$ possesses some principal problems: cosmological equations lead
at physically available initial conditions to not physical solutions \cite{12}
and do not exclude cosmological singularity. Moreover, the presence of the
seconde derivative of the Hubble parameter in cosmological equations leads to
its oscillating behaviour at asymptotics \cite{13}.}. The second restriction
concerns the parameter $q_2$: if $q_2 \neq 0$, the equation for the torsion
function $S_2$ is differential equation of the second order that leads to
oscillating behaviour of the Hubble parameter \cite{6, 8}; by putting $q_2=0$
we will obtain physically necessary consequences. Below we will analyze the
main relations of isotropic cosmology given in \cite{9} in general case without
using any restrictions on indefinite parameters by putting the following
restrictions: $a=0$ and $q_2=0$.

Cosmological equations generalizing Friedmann cosmological equations of GR take
the following form:
\begin{eqnarray}\label{2}
    \frac{k}{R^2} + (H-2S_1)^2 -S_2^2= \nonumber\\
    \frac{1}{{6f_0 Z}}
        \left[
            {\rho  -6 b S_2^2
            + \frac{\alpha }{4} \left( {\rho  - 3p - 12bS_2^2 } \right)^2 }
        \right],
\end{eqnarray}
\begin{eqnarray}\label{3}
    \dot{H}+H^2-2HS_1-2\dot{S}_1 = \nonumber\\
    -\frac{1} {{12f_0 Z}}
        \left[
            \rho  + 3p - \frac{\alpha } {2} \left( {\rho  - 3p - 12bS_2^2 } \right)^2
        \right],
\end{eqnarray}
where $\rho$ is the energy density, $p$ is the pressure, $H=\dot{R}/R $ is the
Hubble parameter (a dot denotes the differentiation with respect to time), the
parameter $\alpha=\frac {f} {3f_0^2}$ ($f>0$) has inverse dimension of energy
density and $Z=1+\alpha\left( \rho - 3p - 12b S_2^2\right)$. The torsion
function $S_1$ is determined by the following way:
\begin{eqnarray}\label{4}
    S_1  = -\frac{\alpha }{4Z} [\dot \rho
    - 3 \dot p + 12f_0 \omega H S_2^2
    \nonumber\\
    -12( {2b - \omega f_0 } ) S_2 \dot S_2],
\end{eqnarray}
where dimensionless parameter $\omega= \frac {2f -q_1} {f} \neq 0$ is
introduced. The torsion function $S_2^2$ satisfies algebraic quadratic
equation, which gives the following root
\begin{eqnarray}\label{5}
 S_{2}^{2}  = \frac{\rho - 3p}{12b} + \frac
{1-(b/2f_0) (1 +  \sqrt{X})} {12b \alpha (1- \omega/4)},
\end{eqnarray}
where $X=1+ \omega (f_0^2/b^2) [1- (b/f_0) - 2(1- \omega /4) \alpha ( \rho +
3p)]$ \footnote{It seems that the second root for $S_2^2$ with opposite sign
before $\sqrt{X}$ in (5) does not lead to physically satisfactory theory.}. In
order to reduce cosmological equations (2)-(3) to closed form we have to
specify the content of HIM and its equation of state. In connection with this
it should be noted that the matter content and its equation of state change
during cosmological evolution and the form of equation of state depends on
coupling of matter with gravitational field. In the case of usual gravitating
matter with energy density $\rho_m$ and pressure $p_m$ coupled minimally with
gravitation the equation of state can be written in usual form:
$p_m=p_m(\rho_m)$ \footnote{Explicit form of equation of state is different at
different stages of cosmological evolution. At present cosmological epoch one
uses the equation of state for dust $p_m=0$, at radiation dominated stage the
equation $p_m=\rho_m/3$ is used.} and the law of energy conservation takes the
form as in GR:
\begin{equation}\label{6}
\dot{\rho}_m+3H\left(\rho_m+ p_m\right)=0.
\end{equation}
We introduce at early stage of cosmological expansion the scalar field $\phi$
with potential $V=V(\phi)$ as component of gravitating matter with the purpose
to investigate inflationary HIM. By minimal coupling with gravitation the
equation for scalar field takes the usual form as in GR:
\begin{equation}\label{7}
\ddot{\phi}+3H\dot{\phi}=-\frac{\partial V}{\partial \phi}.
\end{equation}
Then the total energy density $\rho$ and pressure $p$ are the following:
\begin{eqnarray}\label{8}
\rho=\frac{1}{2}\dot{\phi}^2+V+\rho_m \quad (\rho>0),
p=\frac{1}{2}\dot{\phi}^2-V+p_m.
\end{eqnarray}
Now by using the formula (5) for torsion function $S_2^2$ and eqs. (6)-(8) we
transform the torsion function $S_1$ defined by (4) to the following form:
\begin{eqnarray}\label{9}
S_1  = -\frac{3f_0 \omega \alpha }{4bZ} (HD+E)  ,
 \end{eqnarray}
where
\begin{eqnarray}\label{10}
D = \frac{1}{2} \left(3\frac{d p_m}{d \rho_m}-1 \right) \left(\rho_m+p_m\right)
\nonumber\\
 +\frac{1}{3}\left(\rho_m- 3p_m\right)+\frac{2}{3}\dot{\phi}^2+\frac{4}{3} V
-\frac{b}{6f_0\alpha (1-\omega/4)} \sqrt{X}
\nonumber\\
+\frac{1-\omega (b/2f_0)}{2\sqrt{X}}\Big [\left(3\frac{d p_m}{d\rho_m}+1
\right) \left(\rho_m+p_m\right)+ 4 \dot{\phi}^2 \Big]
\nonumber\\
- \frac{\omega f_0(1-b/f_0)}{b\alpha (1-\omega/4)},
\nonumber\\
E = \left(1+ \frac{1-\omega (f_0/2b)}{\sqrt{X}} \right)\frac{\partial
V}{\partial \phi}\dot{\phi},
\nonumber\\
Z =  \frac{-\omega/4 + (b/2f_0)(1+ \sqrt{X})} {1-\omega/4}.
\end{eqnarray}
By using the formulas for torsion functions we write the cosmological equations
(2)-(3) in the following closed form:
\begin{eqnarray}\label{11}
    \frac{k}{R^2} + \Big [H(1+ \frac{3f_0 \omega \alpha}{2b Z} D) + \frac{3f_0 \omega \alpha}{2b Z}E \Big]^2=
    \nonumber\\
    \frac{1}{6f_0 Z}
       \Big [\rho  + 6(f_0 Z -b) S_2^2
            + \frac{[1-(b/2f_0)(1+\sqrt{X})]^2} {4 \alpha (1-\omega/4)^2 } \Big],
\end{eqnarray}
\begin{eqnarray}\label{12}
   (\dot{H} + H^2)\Big(1+ \frac{3f_0 \omega \alpha}{2b Z} D \Big) +
\nonumber\\
\frac{3f_0 \omega \alpha}{2b Z} \Big [H(\dot{D}-\frac{\dot{Z}}{Z} +E) +\dot{E}-
\frac{\dot{Z}}{Z} E \Big]
\nonumber\\
=-\frac{1}{12f_0 Z} \Big[\rho +3p- \frac{[1-(b/2f_0)(1+\sqrt{X})]^2} {2\alpha
(1-\omega/4)^2 } \Big],
\end{eqnarray}
where the quantities $S_2^2$, $D$, $E$, $Z$ have to be replaced according to
(5) and (10).

By using obtained cosmological equations we will analyze properties of
cosmological solutions at different stages of cosmological evolution.
Simultaneously we will find by what restrictions on indefinite parameters
$\alpha$, $b$ and $\omega$ physical consequences are the most satisfactory and
correspond to observational cosmological data.

At first we will consider the evolution of cosmological models at asymptotics,
when energy densities are small. If the value of dimensional parameter $\omega$
is sufficiently small $|\omega| \ll 1$, the following estimations are valid: $X
\to 1$, $Z \to \frac{b}{f_0}$, $S_1 \to 0$ and the torsion function $S_2^2$
approximately is equal to:
\begin{equation}\label{13}
S_2^2  = \frac{{\rho  - 3p}} {{12b }} + \frac{{1  - b/f_0}} {{12\alpha b}}.
\end{equation}
As a result cosmological equations (2)-(3) take the form of Friedmann
cosmological equations with effective cosmological constant induced by
pseudoscalar torsion function $S_2$:
 \begin{equation}\label{14}
    \frac{k} {{R^2 }} + H^2  = \frac{1} {{6f_0 }}\left[ \rho (f_0/b) + \frac{1}{4} \alpha^{-1}(1 - b/f_0)^2 (f_0/b)
    \right],
\end{equation}
\begin{equation}\label{15}
    \dot H + H^2  =  - \frac{1} {{12f_0 }}\left[ (\rho + 3p)(f_0/b) - \frac{1}{2} \alpha^{-1}(1 - b/f_0)^2 (f_0/b)\right].
\end{equation}
By certain values of parameters $\alpha$ and $b$ physical consequences of eqs.
(14)-(15) for flat model ($k=0$) without introducing of DE and DM are identical
to that of standard $\Lambda CDM$-model \cite{7}. In fact, the energy density
$\rho$ in eqs. (14)-(15) corresponds to total energy density of physical matter
in the Universe, which consists of barionic matter, relic radiation, neutrino
etc. If dark matter does not exist, the value of $\rho$ is approximately equal
to energy density of barionic matter, because the contribution of other named
components is sufficiently small, that leads to certain estimation of the
parameter $b$ ($b \approx f_0/6$). Then by taking into account cosmological
data concerning the dark energy we obtain the estimation of the parameter
$\alpha$ \cite{7} \footnote{In the case if some DM exists, the value of $b$
increases being less than $f_0$ that leads to corresponding change of
$\alpha$.}. Because the cosmological equations (14)-(15) were obtained in
zeroth approximation with respect to small parameter $\omega$, the following
question appears: what are quantitative limits of applicability of these
equations? In order to find estimation of parameter $\omega$, we will analyze
the behaviour of cosmological solutions in the beginning of cosmological
expansion.

First of all important physical consequences follow from formula (5) for
$S_2^2$-function. If the parameter $\omega$ is positive ($0<\omega \ll 1$),
because the value of $X$ can not be negative we obtain principal constraint for
admissible energy densities: $X=1+ \omega (f_0^2/b^2) [1- (b/f_0) - 2(1- \omega
/4) \alpha ( \rho + 3p)]\ge 0$ or by taking into account smallness of $\omega$
the following relation:
\begin{equation}\label{16}
X = 1 - 2 (f_0^2/b^2)\omega  \alpha ( \rho + 3p)\ge 0.
\end{equation}
In the case of systems filled with usual matter with energy density $\rho_m$
without scalar fields the equality defined by (16) determines a limiting
(maximum) energy density. When energy density $\rho_m$ is comparable with
$\rho_{max}$, the gravitational interaction has the character of repulsion
ensuring the regularity of such systems \footnote{In the frame of PGTG with
gravitational Lagrangian (1) the conclusion about existence of limiting mass
density was obtained in the case of HIM with the only torsion function $S_1$ in
\cite{14} (see also \cite{11}). Later the hypothesis about existence in the
nature of limiting mass density equal to Planckian one was discussed in
\cite{15}.}. In the frame of HIM without dark matter the value of $\alpha^{-1}$
is comparable with average energy density in the Universe at present epoch. The
order of limiting energy density $\rho_{max}$ is determined by the value of
$(\omega \alpha)^{-1}$. If the value of $\rho_{max}$ is comparable or greater
than energy density of quark-gluon plasma (but less than the Planckian one), we
obtain that dimensionless parameter $\omega$ is extremely small. In connection
with this the cosmological equations (14)-(15) are very good approximation at
least for matter dominated stage of cosmological evolution. In the case of
systems including also scalar fields, for which energy density and pressure are
defined by (8), the relation (16) determines in space of matter parameters
$(\rho_m, \phi, \dot{\phi})$ domain of their admissible values. This domain is
limited by surface $L$ defined by $X=1 - 2 (f_0^2/b^2)\omega  \alpha ( \rho_m +
3p_m + 2\dot{\phi}^2- 2V)=0$. Moreover, it is necessary to take into account
that additional restriction on admissible values of matter parameters $(\rho_m,
\phi, \dot{\phi})$ follows from positivity of expression (5) for $S_2^2$.

Now we will analyze the behaviour of cosmological solutions near the limiting
energy density or limiting surface $L$, where $X \ll 1$. With this purpose we
consider the expression of the Hubble parameter $H$ following from cosmological
equation (11):
\begin{eqnarray}\label{17}
H_{\pm}=\Bigg[-\frac{3f_0 \omega \alpha}{2b Z}E \pm \Big(\frac{1}{6f_0 Z}
       \Big [\rho  + 6(f_0 Z -b) S_2^2
       \nonumber\\
            + \frac{[1-(b/2f_0)(1+\sqrt{X})]^2} {4 \alpha (1-\omega/4)^2 }
            \Big]
            -\frac{k}{R^2}\Big)^{1/2} \Bigg]\Big(1+ \frac{3f_0 \omega \alpha}{2b Z}
D\Big)^{-1}
\end{eqnarray}
Similarly to isotropic cosmology based on HIM with the only torsion function
\cite{2,3,4}, at asymptotics where energy densities are sufficiently small,
$H_{-}$-solutions correspond to cosmological compression and $H_{+}$-solution
-- to cosmological expansion, and the transition from $H_{-}$ to
$H_{+}$-solution take place by reaching the limiting energy density or limiting
surface $L$ (see below) \footnote{In the case of HIM of closed type $(k=+1)$
there are also solutions without reaching the limiting surface $L$ (see
\cite{2}).}. In order to obtain some physical characteristics of such
transitions the formula (17) for $H_{\pm}$ can be simplified by taking into
account the smallness of parameter $\omega$ ($\omega \ll 1$) and also that at
considering extreme conditions: $\alpha^{-1} \ll \rho$, $X\ll 1$, $\rho \sim
(\omega \alpha)^{-1}$. As a result expressions of $D$, $E$, $Z$ and $S_2^2$ are
simplified:
\begin{eqnarray}\label{18}
 D=\frac{1}{2}
\left(3\frac{d p_m}{d \rho_m}-1 \right) \left(\rho_m+p_m\right)
+\frac{1}{3}\left(\rho_m- 3p_m\right)+\frac{2}{3}\dot{\phi}^2
\nonumber\\
+\frac{4}{3} V+\frac{1}{2\sqrt{X}}\Big [\left(3\frac{d p_m}{d \rho_m}+1 \right)
\left(\rho_m+p_m\right)+ 4 \dot{\phi}^2 \Big],
\nonumber\\
E = \left(1+ \frac{1}{\sqrt{X}} \right)\frac{\partial V}{\partial
\phi}\dot{\phi}, Z=\frac{b}{2f_0} (1+ \sqrt{X}),
\nonumber\\
 S_{2}^{2}  = \frac{\rho - 3p}{12b} + \frac
{1-(b/2f_0) (1 +  \sqrt{X})} {12b \alpha},
\end{eqnarray}
where the quantity $X$ is defined by (16). By taking into account terms linear
with respect to $\sqrt{X}$ the expression (17) for $H_{\pm}$ takes the
following form:
\begin{eqnarray}\label{19}
H_{\pm}=\Bigg[-2 \frac{\partial V}{\partial \phi}\dot{\phi} +\sqrt{X}\Big[-2
\frac{\partial V}{\partial \phi}\dot{\phi}
\nonumber\\
\pm \frac{2b^2}{3f_0^2 \omega \alpha}
      \Big(\frac{1}{4b}(\rho_m+p_m\ + \dot{\phi}^2)-\frac{k}{R^2}\Big)^{1/2}\Big]\Bigg]
       \nonumber\\
        \Bigg[ \Big [\left(3\frac{d p_m}{d\rho_m}+1 \right)
\left(\rho_m+p_m\right)+ 4 \dot {\phi}^2 \Big]
\nonumber\\
+ {\sqrt{X}}\Big [\frac{2b^2}{3f_0^2 \omega \alpha}+  \Big( (3\frac{d
p_m}{d\rho_m}-1) \left(\rho_m+p_m\right)
\nonumber\\
+\frac{2}{3}\left(\rho_m- 3p_m\right)+\frac{4}{3}\dot {\phi}^2+\frac{4}{3} V
\Big)\Big] \Bigg]^{-1}.
\end{eqnarray}
In the case of models filled with usual matter without scalar fields in linear
approximation with respect to $\sqrt{X}$ the Hubble parameter and its time
derivative take the following form:
\begin{eqnarray}\label{20}
H_{\pm}=\pm \frac{2b^2}{3f_0^2 \omega \alpha} \frac{\sqrt{X} [(1/4b)(\rho_m
+p_m) - (k/R^2)]^{1/2}}{(3\frac{d p_m}{d\rho_m}+1 ) (\rho_m+p_m)^{1/2}}
\nonumber\\
\dot{H}=\frac{4b^2}{3f_0^2\omega \alpha } \frac{(1/4b)(\rho_m +p_m) -
(k/R^2)}{(3\frac{d p_m}{d\rho_m}+1 ) (\rho_m+p_m)^{1/2}}.
\end{eqnarray}
By reaching a limiting energy density ($X=0$) the Hubble parameter vanishes and
the value of its time derivative is the same for $H_{-}$- and $H_{+}$-solution
and it is positive that corresponds to a bounce. By using obtained expression
for the Hubble parameter it is easy to show that by given equation of state
$p_m=p_m(\rho_m)$ the evolution of scale factor $R(t)$ near a bounce take the
following form: $R(t)=R_{min}+ r_1 t^2+...$, where $t=0$ corresponds to a
bounce, $R_{min}$ is minimum value of $R$ depending on limiting energy density
and given equation of state, the value of $r_1>0$ is expressed by $\dot{H}$ at
a bounce. In the case of models including also scalar fields the Hubble
parameter does not vanish by reaching a limiting surface $L$ and according to
(19) its value on surface $L$ is:
\begin{eqnarray}\label{21}
H_L=\frac{-2 \frac{\partial V}{\partial \phi}\dot{\phi}}{(3\frac{d
p_m}{d\rho_m}+1) \left(\rho_m+p_m\right)+4 \dot{\phi}^2}.
\end{eqnarray}
The bounce in this case takes place in points of extremum surface in space of
matter parameters $(\rho_m, \phi, \dot{\phi})$, equation of which we obtain by
setting $H=0$ in cosmological equation (11). Cosmological solutions can be
found by numerical integration of eqs. (12), (6) and (7) by choosing initial
conditions on extremum surface. Similarly to inflationary cosmological models
built in the frame of isotropic cosmology with the only torsion function (see
\cite{2,3}), if initial value of scalar field is sufficiently large we obtain
regular inflationary solution containing transition stage from compression to
expansion, inflationary stage with slow-rolling behaviour of scalar field and
post-inflationary stage with oscillating scalar field. Similarly to our works
\cite{2,3} regular Big Bang scenario can be built on the base of such
inflationary cosmological models. All cosmological solutions are regular with
respect to energy density, the scale factor $R$ and the Hubble parameter $H$.
Unlike HIM with the only torsion function $S_1$, in the case of considered HIM
with two torsion functions the torsion does not diverge by reaching limiting
surface $L$ (or limiting energy density): the torsion function $S_2$ is regular
and the torsion function $S_1$ undergoes a finite jump by reaching the surface
$L$ (or limiting energy density).

The solution of cosmological problems in the frame of PGTG presented above is
achieved by virtue of the change of gravitational interaction in comparison
with GR and Newton's theory of gravity at cosmological scale. These changes are
provoked by more complicated structure of physical spacetime, namely by
spacetime torsion. In connection with this the following question appears: what
situation takes place in the case of gravitating systems at other spatial
scales (galaxies, stars, solar system), what possible role the torsion plays in
these systems? First of all, the torsion should be important in gravitating
systems at astrophysical scale (galaxies and their accumulation), by
investigation of which the notion of dark matter was introduced more than 80
years ago. It should be noted that the investigation of such systems in the
frame of PGTG is difficult mathematical problem, because the system of
gravitational equations of PGTG in the case of non-homogeneous gravitating
systems, including spherically symmetric and axially symmetric objects, is very
complicated system of differential equations, and so far the DM-problem at
astrophysical scale is not solved. The conclusion obtained in this paper about
possible existence of limiting energy density and gravitational repulsion
effect at extreme conditions can be of principal meaning for theory of massive
superdense stars, where gravitational repulsion effect has to prevent a
collapse. Concerning the solar system, usual gravitational effects including
relativistic corrections in this case can be obtained in the frame of PGTG,
because the vacuum Schwarzschild solution for the metrics with vanishing
torsion is exact solution and does not depend on values of indefinite
parameters of gravitational Lagrangian (1) \cite{16}. However, from physical
point of view this solution has certain limits of its applicability, because
the physical spacetime in the vacuum in the frame of accelerating Universe is
de Sitter spacetime with non-vanishing torsion \cite{9}.

\end{document}